# The Miniature Optical Transmitter and Transceiver For the High-Luminosity LHC (HL-LHC) experiments

**Chonghan Liu**[a*], **Xiandong Zhao**[a], **Jinghong Chen**[b], **Binwei Deng**[a,c], **Datao Gong**[a], **Di Guo**[a,d], **Deping Huang**[b], **Suen Hou**[e], **Xiaoting Li**[a,f], **Futian Liang**[a,d], **Gang Liu**[a,g], **Tiankuan Liu**[a], **Ping-Kun Teng**[e], **Annie C. Xiang**[a], **Jingbo Ye**[a]

[a] *Department of Physics, Southern Methodist University, Dallas TX 75275 USA*

[b] *Department of Electrical Engineering, Southern Methodist University, Dallas, TX 75275, USA*

[c] *Hubei Polytechnic University, Huangshi, Hubei 435003, P. R. China*

[d] *Department of Modern Physics, University of Science and Technology of China, Hefei, Anhui 230026, P. R. China*

[e] *Institute of Physics, Academia Sinica, Nangang 11529, Taipei, Taiwan (Suen, PK, Da-shung)*

[f] *School of Physics, Central China Normal University, Wuhan Hubei 430079 China*

[g] *Institute of High Energy Physics, Chinese Academy of Sciences, Beijing 100049, P. R. China*
   *E-mail*: kentl@smu.edu

ABSTRACT: We present the design and test results of the Miniature optical Transmitter (MTx) and Transceiver (MTRx) for the high luminosity LHC (HL-LHC) experiments. MTx and MTRx are Transmitter Optical Subassembly (TOSA) and Receiver Optical Subassembly (ROSA) based. There are two major developments: the Vertical Cavity Surface Emitting Laser (VCSEL) driver ASIC LOCld and the mechanical latch that provides the connection to fibers. In this paper, we concentrate on the justification of this work, the design of the latch and the test results of these two modules with a Commercial Off-The-Shelf (COTS) VCSEL driver.

**Contents**

**1. Introduction**

Optical links have been extensively used in high energy physics experiments. Optical transmitters and optical transceivers are the key components to build optical links. For many high energy physics experiments, all components, including optical transmitters and optical receivers, mounted on the high energy physics detector must operate in harsh radiation environment. The radiation tolerance of various commercial optical transmitters and optical transceivers has been studied [1] and found to be hard to meet the requirements of many high energy physics experiments and thus custom-made optical transmitters and optical transceivers must be developed. One of such high energy physics experiments is the high luminosity LHC (HL-LHC) upgrade. For example, in the update of the ATLAS liquid argon calorimeter (LAr) trigger upgrade [2], a board that reads out up to 320 detector channels needs to be developed. This board, called the LAr Trigger Digitizer Board (LTDB), will be placed in the existing front-end-crate (FEC) and be radiation tolerant to operations after LHC's upgrades. The analog signals of this board will be processed, digitized and transmitted off the detector at a bandwidth of 204.8 Gb/s carried by 40 fibers. The Miniature optical Transmitter (MTx) will be a dual-channel module that operates at 5.12 Gb/s per channel. It will be board-mounted and fit into the existing mechanical constraints of all the boards in the FEC. The module needs to have a height of 6 mm and be small in size to allow 20 MTx on the LTDB. Given the size of the LTDB and the serializer ASIC that provides data to MTx, and the reliability concerns of the Vertical Cavity Surface Emitting Laser (VCSEL), we choose to use Transmitter Optical Subassembly (TOSA) and Receiver Optical Subassembly (ROSA) in MTx and MTRx, and develop a custom latch to connect them to fiber and to meet the 6 mm height requirement. The idea of a board-mount optical transceiver with the serializer in its belly is taken from the Small-Factor Versatile Transceiver (SF-VTRx) of the Versatile Link common project [3]. Though MTx and MTRx are designed for the ATLAS LAr trigger upgrade, they can also be used in other high energy physics experiments which require radiation tolerance and low profile.

## 2. Design

### 2.1 Latch

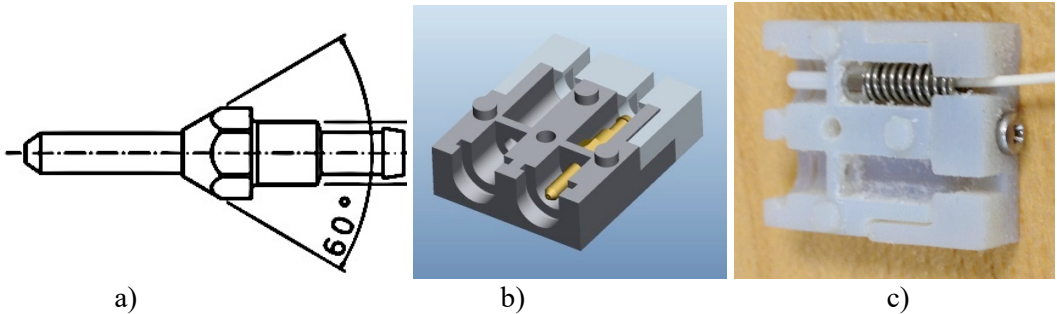

a)                  b)                  c)

Figure 1. a) The standard LC ferrule with flange. d) 3D representation of the latch. c) 3D printed latch prototype.

The heights of a regular LC latch and an LC connector are more than 6 mm. We choose the fiber with a standard ferrule as shown in Figure 1(a). The fiber can be ordered from fiber vendors. We keep the spring loading mechanism to maintain fiber-TOSA/ROSA connection, but we remove the outer case of the standard LC connector and latching system to reduce the height.

    The latch is designed in two pieces. One piece is used to lock the TOSAs on the Printed Circuit Board (PCB) by three pins and one screw. The other piece loads the fiber with a spring and attaches to the first piece with a screw. The 3D representation of the latch is shown in Figure 1(b). The optical coupling is maintained by the ferrule on the fiber and TOSA/ ROSA package. The prototype was 3D-printed out as shown in Figure 1(c). The latch will finally be produced by using injection molding with irradiation tolerant material.

### 2.2 Printed Circuit Board

#### 2.2.1 MTx

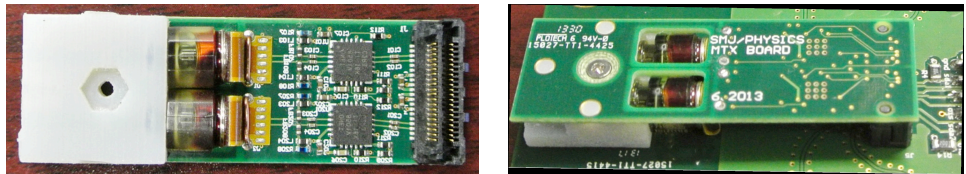

Figure 2. a) MTx PCB board with a latch    b) MTx module plugged on a carrier board.

The pictures of an MTx PCB with a latch and an MTx module plugged on a carrier board are shown in Figure 2. The optical components are VCSEL TOSAs and the VCSEL drivers are ONET8501V from Texas Instrument. The VCSEL used here has been tested to be radiation tolerant [4]. ONET8501V is radiation soft. A radiation tolerant VCSEL driver ASIC called LOCld is being developed [5]. Once LOCld is tested, it will replace the COTS VCSEL driver. The electrical interface uses an LSHM connector (LSHM-120-2.5-L-DV-A-N-T-R-C from Samtec) which has a speed of 7 GHz for differential pair signals. Two rectangular holes are opened on PCB board to accommodate the body of the TOSA/ROSA. All the parts are arranged on only one side of the PCB board to reduce the height of the whole module.

### 2.2.2 MTRx

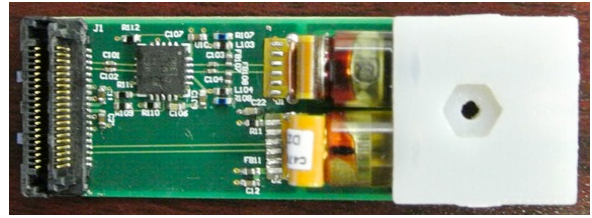

Figure 3. MTRx PCB board with latch

A picture of MTRx is shown in Figure 3. MTRx re-uses most design of MTx. It keeps one transmitting channel and replace another channel with a GigaBit Transimpedance Amplifier (GBTIA) loaded ROSA [6]. GBTIA is designed to work at 5 Gbps.

### 2.2.3 Comparison with SFP+

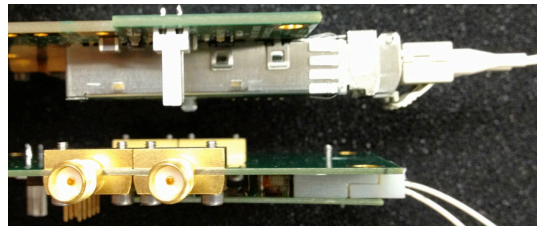

Figure 4. MTx vs. SFP+ plugged on board

An MTx module and a commercial SFP+ module are compared in Figure 4. The size of MTx/MTRx is 45 mm (length) × 15 mm (width) × 6 mm (height) and the size of an SFP+ transceiver (cage) is 48.7 mm (length) × 14.5 mm (width) × 9.7 mm (height). An SFP+ module can be as high as 14 mm when the metal cage and the LC connector are counted. MTx/MTRx not only has a smaller form factor but also less material than a commercial SFP+ module.

## 3. Testing

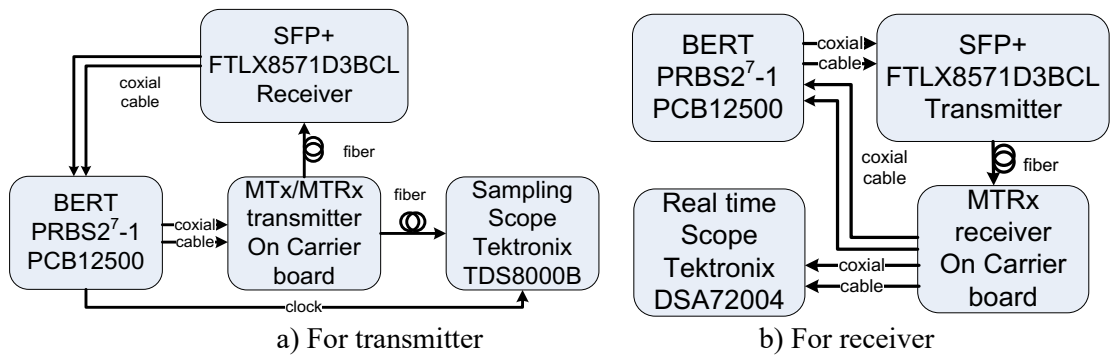

Figure 5. Test benches for eye diagram test and BER test.

The optical test bench is shown in Figure 5. All the parts for the MTx and the module design are targeted at 10 Gbps signal transmission. For the MTRx only the ROSA loaded with GBTIA works at no more than 5 Gbps. Therefore the transmitter side was tested at both 5 Gbps and 10 Gbps. The receiver side was tested only at 5 Gbps.

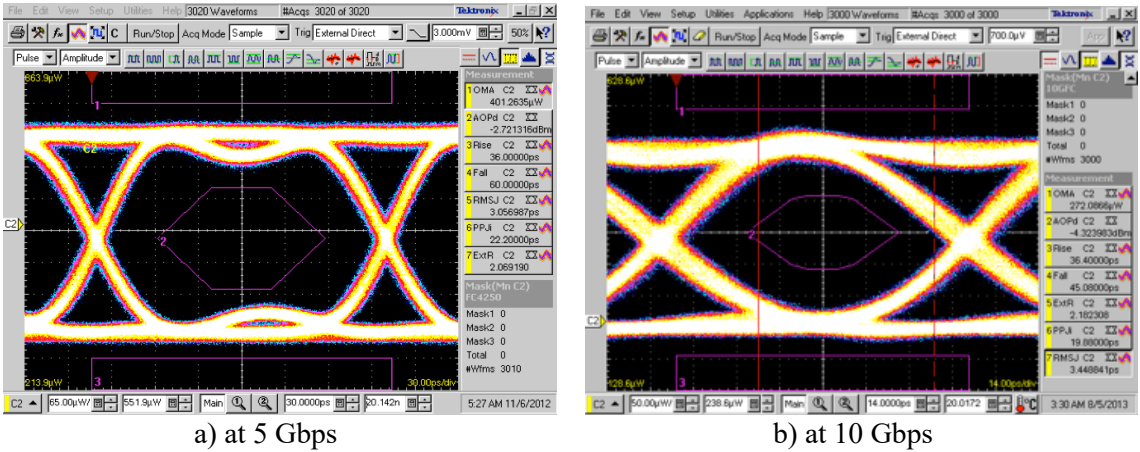

a) at 5 Gbps          b) at 10 Gbps

Figure 6. Typical transmitter Optical Eye Diagrams

Figure 6 shows the typical eye diagrams of the transmitter at 5 Gbps and 10 Gbps, respectively. The 10 Gbps eye diagrams passed the eye mask of the standard 10G fiber channel when the bias current was set at 8 mA and the modulation current was programmed from 2.5 mA to 14 mA (the corresponding optical modulation amplitude (OMA) was from -10.25 dBm to -3.13 dBm). Figure 7 shows the typical eye diagram of the receiver at 5 Gbps.

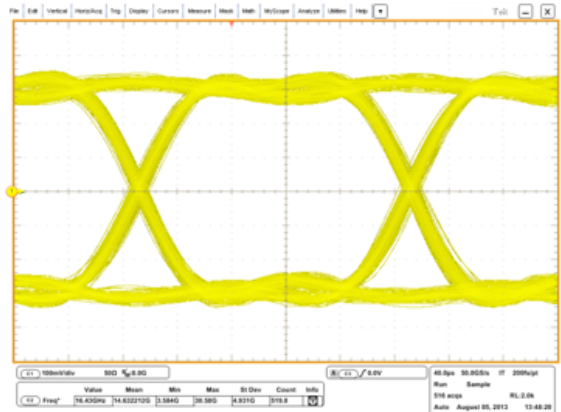

Figure 7. Typical receiver electrical eye diagram at 5 Gbps

Bit Error Rate (BER) tests were performed at 10 Gbps data rates on the transmitter where the receiver is a Finisar SFP+ transceiver (FTLX8571D3BCL). When the OMA is from -10.25 dBm to -3.13 dBm the BERs are less than $10^{-12}$ at a 95% confidence level. BER test was also performed at 5 Gbps on the receiver side where the transmitter is the same Finisar SFP+ module. The BER is less than $10^{-12}$ at a 95% confidence level.

## 4. Conclusion

TOSA and ROSA based miniature optical dual channel transmitter MTx and optical transceiver MTRx meet design specifications. With a COTS VCSEL driver, the transmitter is tested up to 10 Gbps. With LOCld, MTx and MTRx are expected to be radiation tolerant for ATLAS LAr trigger upgrade and potentially for other HL-LHC experiments. A pre-production of MTx will be carried out for the LTDB demonstrator and a life test will be performed to assess the module's reliability.

## Acknowledgments

We thank Csaba Soos of CERN for many inspiring discussions. This project has been supported by US ATLAS Upgrade Funds, the US Department of Energy Collider Detector Research and Development (CDRD) and the National Science Council in Taiwan.